\documentclass[%
 aip,
 jmp,%
 amsmath,amssymb,
 reprint,%
]{revtex4-1}

\usepackage{graphicx}
\usepackage{dcolumn}
\usepackage{bm}
\usepackage{hyperref}
\usepackage{boxhandler}
\usepackage{float}

\let\oldhat\hat
\renewcommand{\hat}[1]{\oldhat{\mathbf{#1}}}
\begin{document}


\title{Resolution of spin Hall and anisotropic magnetoresistance in Pt/EuO$_{1-x}$}

\author{Kingshuk Mallick}
	\affiliation{Physics Department, Indian Institute of Science, Bangalore - 560012, India}
\author{Aditya A. Wagh}%
	\affiliation{Physics Department, Indian Institute of Science, Bangalore - 560012, India}%
\author{Adrian Ionescu}%
	\affiliation{Cavendish Laboratory, Physics Department, University of Cambridge, Cambridge CB3 0HE, United Kingdom} 
\author{Crispin H.W. Barnes}%
	\affiliation{Cavendish Laboratory, Physics Department, University of Cambridge, Cambridge CB3 0HE, United Kingdom} 
\author{P.S.Anil Kumar}%
 \email{anil@iisc.ac.in}
\affiliation{Physics Department, Indian Institute of Science, Bangalore - 560012, India}%
\date{\today}
\begin{abstract}
We report on the angular and field dependence of the magnetoresistance (MR) in bilayers of Pt/EuO$_{1-x}$ thin films, measured in both in-plane and out-of-plane geometries at different temperatures (\textit{T}). Presence of oxygen vacancies manifested by a metal-insulator transition as well as a high-\textit{T} ferromagnet to paramagnet transition (T$_P$) were observed in the bilayers. 
The Anisotropic Magnetoresistance (AMR) could be extracted in the entire \textit{T}-range, even above T$_P$, exhibiting two sign crossovers. We attribute its \textit{T}-evolution to the rotation of easy axis direction from a high-\textit{T} out-of-plane to a low-\textit{T} in-plane orientation. In addition, considering MR contributions from the films' (111) texture and interface, we identify a \textit{T}-window wherein the spin Hall effect induced spin Hall magnetoresistance (SMR) could be extracted. 
\end{abstract}

\maketitle

%

EuO has a rocksalt structure(a=0.5144 nm) whose ferromagnetic response below its Curie temperature (T$_c$) of 69 K is due to the half-filled 4$f$ Eu$^{2+}$ orbital\cite{passell1976neutron}. Oxygen deficient EuO, i.e. EuO$_{1-x}$ is intrinsically electron doped and undergoes a simultaneous ferromagnetic and insulating-conducting phase transition, while the conduction electrons become nearly 100$\%$ spin polarized\cite{steeneken2002exchange}. Such doping is also known to enhance the bulk $T_{c}$ to above 140 K\cite{steeneken2002exchange}. However, very little has been explored in this material with respect to its magnetotransport properties\citep{rimal2018interface}. 

The variation of electrical resistance on application of magnetic fields is broadly termed as magnetoresistance (MR). 
In bilayers of a heavy metal (HM) and ferromagnetic insulator (FI) a new type of MR has been discovered that utilizes the large spin-orbit coupling of the HM to generate (detect) spin currents via the direct (inverse) spin Hall effect, SHE (ISHE). This phenomena is referred to as the spin Hall magnetoresistance (SMR)\cite{althammer2013quantitative} observed first in a Pt/YIG bilayer. It has proven instrumental in quantifying the spin mixing conductance (G$_{mix}$) of HM/FI bilayers\cite{wang2015spin,velez2018spin}, which influences several spin transport phenomena such as the spin Seebeck effect\cite{mallick2019enhanced,mallick2019role}. 

In a HM/Ferromagnet (FM) bilayer, if the FM is conducting or if proximity induced magnetism is developed in the the HM, as in Pt and Pd, then along with the SMR, AMR is also detected in magnetic field orientation dependent MR (or angle dependent MR, ADMR) and magnetic field amplitude dependent MR (or field dependent MR, FDMR)\cite{kim2016spin,zhou2018spin}. AMR  originates from the spin-orbit coupling in FM\cite{mcguire1975anisotropic} and hence is sensitive to the charge, orbital, spin and lattice degrees of freedom. For instance, it has been used to probe systems undergoing simultaneous metal-insulator and ferromagnet-paramagnet phase transition like in manganites\cite{alagoz2015mechanism}. In particular, AMR has been shown to be a useful tool to probe bulk magnetic domain configuration, exhibiting a sign change or crossover as a function of \textit{T} under a spin reorientation transition\cite{kumar2019magnetotransport,miyakozawa2016temperature}. 


Here we report on the observation of SMR and AMR in the bilayer of Pt and EuO$_{1-x}$. The sign of the AMR changes twice in the \textit{T}-range under study and remains finite even above T$_{P}$. The AMR results are compared with the bulk magnetization measurements to understand how the magnetization state influences the resistance. In contrast, the SMR is negative and vanishes above T$_P$.

The nominal textured polycrystalline sample structure used for the experiments is depicted in Fig. \ref{fig:xrd-vsm}(a). A field cooled \textit{M-T} for the sample exhibits two dome-like features (see supplementary Fig. 1). The low-\textit{T} feature observed at 65 K (T$_{EuO}$), is close to the T$_c$ observed in bulk stoichiometric EuO ($\sim$69 K) and another at 144 K (T$_P$) corresponds to the extended T$_{c}$. 
The soft ferromagnetic nature and in-plane easy axis at 10 K is evident from the \textit{M-H} hysteresis loops depicted in Fig.\ref{fig:xrd-vsm}(c), determined using SQUID magnetometery. 
Futher details of the sample deposition conditions and additional structural and magnetic characterization can be found in \cite{mallick2019role}.

\begin{figure}[t]
	\begin{center}
		\includegraphics[width=0.4\textwidth]{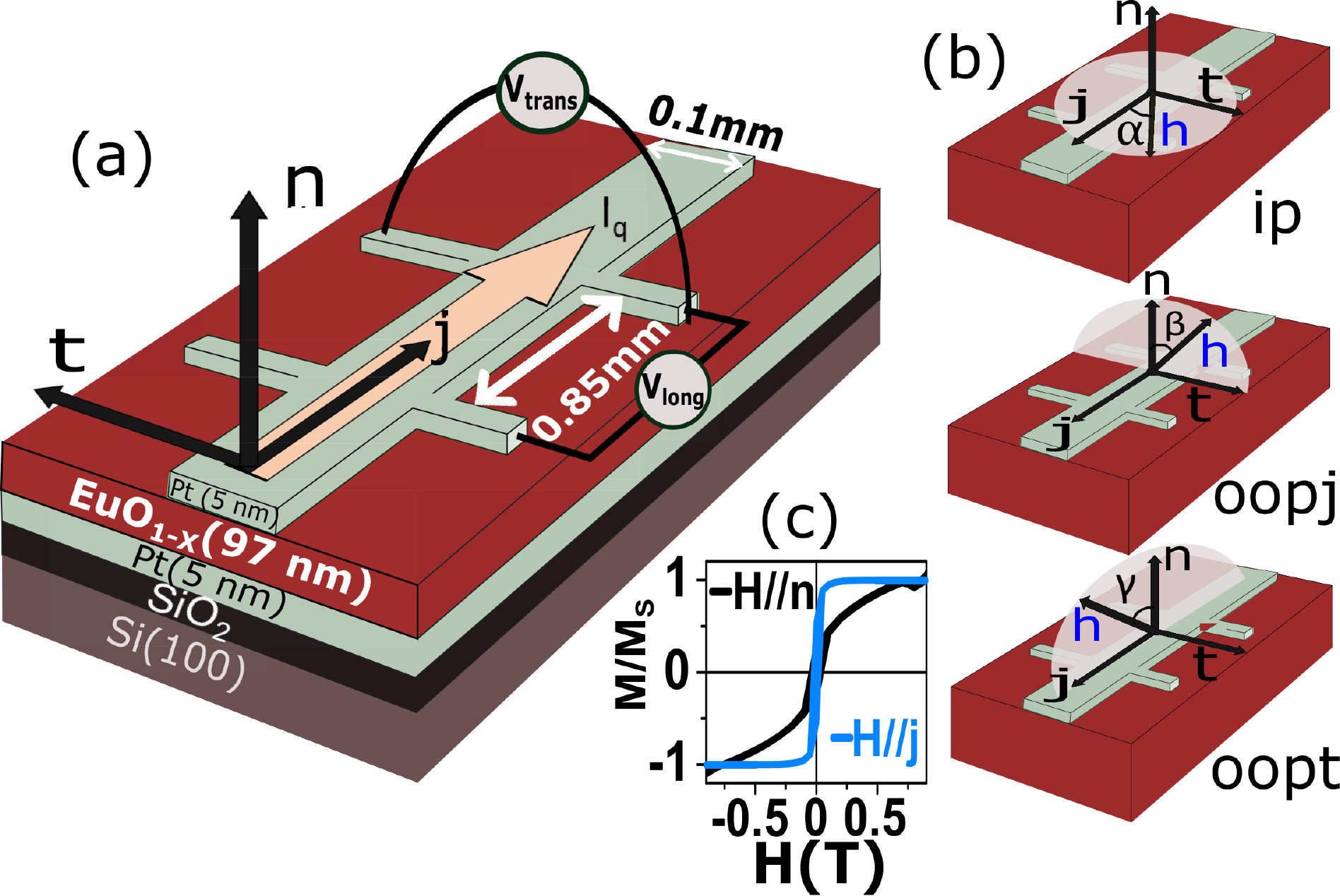}
		\small{\caption{(a) Schematic of the sample stack, Hall bar and the measurement configuration. The adopted coordinate system defined by \textbf{j}, \textbf{t} and \textbf{n} are labeled. (b) Different configuration adopted in ADMR measurements.
		(c) \textit{M-H} hysteresis loops at 10 K with field applied along \textbf{j} and \textbf{n} highlighting the ip easy axis direction.\label{fig:xrd-vsm}}}
	\end{center}	
\end{figure}

Resistivity determination in bare EuO$_{1-x}$ is tedious owing to the difficulty in getting proper ohmic contacts\cite{altendorf2011oxygen}. 
IN our previous study\cite{mallick2019role}, we demonstrated that an estimate of the resistivity of our EuO$_{1-x}$ layer can be made from the measured trilayer resistance of the entire stack, by assuming three resistances corresponding to the three layers connected in parallel (see Fig. 2 in supplementary for further details). Comparing to other reports on EuO$_{1-x}$\cite{steeneken2012new}, we find that this approach captures the main features of the \textit{T}-dependent resistivity, particularly the MIT at T$_{EuO}$, as well as the low-\textit{T} order of magnitude, reasonably well. 

For magnetotransport experiments we pattern the top Pt layer into Hall bar structures using standard photolithography and argon ion milling as illustrated in Fig. \ref{fig:xrd-vsm}(a). The four probe longitudinal resistance, $R_{long}$ (hereafter just \textit{R}) was calculated from the measured voltage $V_{long}$, using lock-in technique under an applied AC current of frequency 333 Hz. For ADMR, we rotate a constant magnetic field of 0.25 T along \textit{jt}, \textit{nt} and \textit{nj} planes with angles $\alpha$ (ip), $\beta$ (oopt) and $\gamma$ (oopj) with respect to \textbf{j}-, \textbf{n}- and \textbf{n}-axes, respectively (see Fig. \ref{fig:xrd-vsm}(b)). FDMR is studied in magnetic fields upto 0.35 T, applied along the two orthogonal axes, in each of the three configurations, ip, oopt and oopj. The maximum applied fields are larger than the coercive fields for both in-plane (ip) and out-of-plane (oop) orientation and hence \textbf{m} can be assumed to follow \textit{\textbf{M}}. To determine the \textit{T}-dependence, the sample was thermally anchored in a modified closed cycle cryostat and cooled from 300 K to 10 K and subsequently ADMR and FDMR was recorded by warming to the desired-\textit{T}. First, we will discuss the observed MR in the $\gamma$ scan which is sensitive to the AMR arising from current flow in the conducting EuO$_{1-x}$. 

\begin{figure}[ht]
	\begin{center}
		\includegraphics[width=0.5\textwidth]{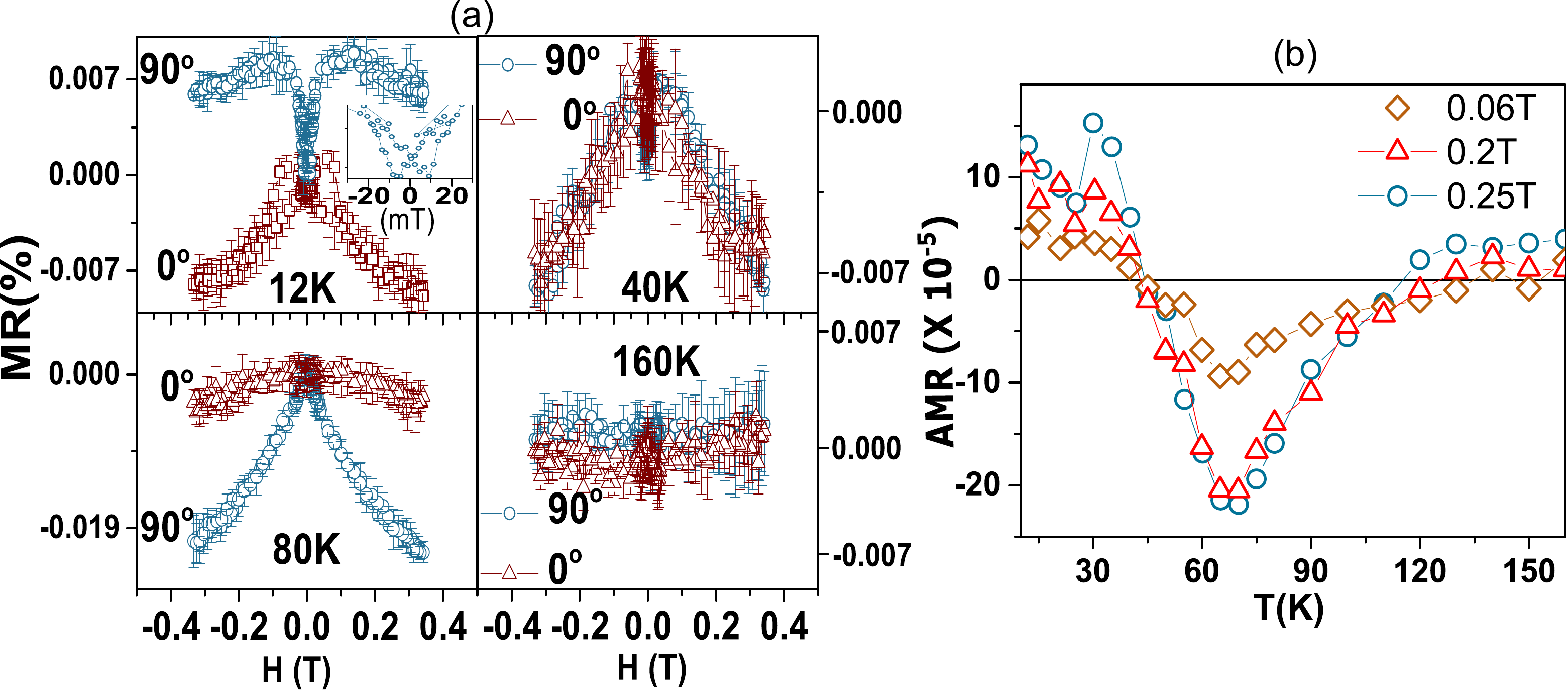}
		\small{\caption{(a)  $R(H)$ at different \textit{T} for \textit{H} oriented parallel ($\gamma=0^o$) and perpendicular ($\gamma=90^o$) to \textbf{n}. Inset depicts the expanded view near the origin showing the hysteretic switching of \textit{R}. (b) \textit{T}-dependence of the scaled AMR. The 0.25 T data corresponds to ADMR results whereas data for other two \textbf{\textit{H}} values are extracted from FDMR. \label{fig:oopt}}}          
	\end{center}	
\end{figure}

The main results are highlighted in Fig. \ref{fig:oopt}. First, we verified the existence of a MR effect by recording the variation in \textit{R} as a function of \textit{H} ($R(H)$). Figure \ref{fig:oopt}(a) shows the $MR$ percentage ($=100\times(R(H=0)-R(H))/R(H=0)$) obtained at different-\textit{T} with \textit{H} oriented along \textbf{j} ($\gamma = 90^o$) and \textbf{n} ($\gamma = 0^o$). We observe characteristic hysteretic switching behavior of \textit{R} at the coercive fields (see inset of Fig. \ref{fig:oopt}(a) at 12 K) for both \textit{H} orientations at low \textit{T}, however, it becomes progressively less evident with an increase in \textit{T}. This resembles the isothermal $M(H)$ loops, which also shows vanishing coercivity at high-\textit{T}\cite{mallick2019role}. 
The non-saturating nature of $MR$, within the accessible range of \textbf{H}, particularly in the \textit{ip} direction, can be explained based on the CMR effect, that is known to contribute a significant negative $MR$, isotropic in \textbf{H}, around the MIT for EuO$_{1-x}$\cite{steeneken2012new}. According to AMR definition, a measure of the AMR magnitude, $\Delta R_{AMR}$, is given by, $\Delta R_{AMR}=R(H||j)-R(H||n)$. We determine it at two \textit{H} values, 0.06 T ($H\sim H_{c}$) and 0.2 T ($H\gg H_{c}$), where H$_{c}$ is out-of-plane coercive field at 10 K. It is taken to be positive when $R(H||j)$ $>$ $R(H||n)$. The AMR ratio, $\Delta R_{AMR}/R(H||n)$ obtained at different-\textit{T} are presented in Fig. \ref{fig:oopt}(b). Three stricking features that emerge from the \textit{T}-dependence are, (i) the AMR is non-zero even above T$_P$, (ii) the sign-change or crossover happens twice in the \textit{T}-range and (iii) it exhibits a minimum at T$_{EuO}$. These characteristics are reproduced by the AMR ratio calculated from ADMR experiments as well (0.25 T data in Fig.  \ref{fig:oopt}(b)). Some representative ADMR results at different-\textit{T} are shown in supplementary Fig. 3, which undoubtedly captures the sign change as 90$^o$ phase shifts in the $\gamma$ dependence. The peak feature in EuO$_{1-x}$ suggests presence of competing interactions which we will now discuss by comparing our results with bulk magnetometry measurements performed using a SQUID magnetometer. 

\begin{figure*}[ht]
	\begin{center}
		\includegraphics[width=0.99\textwidth]{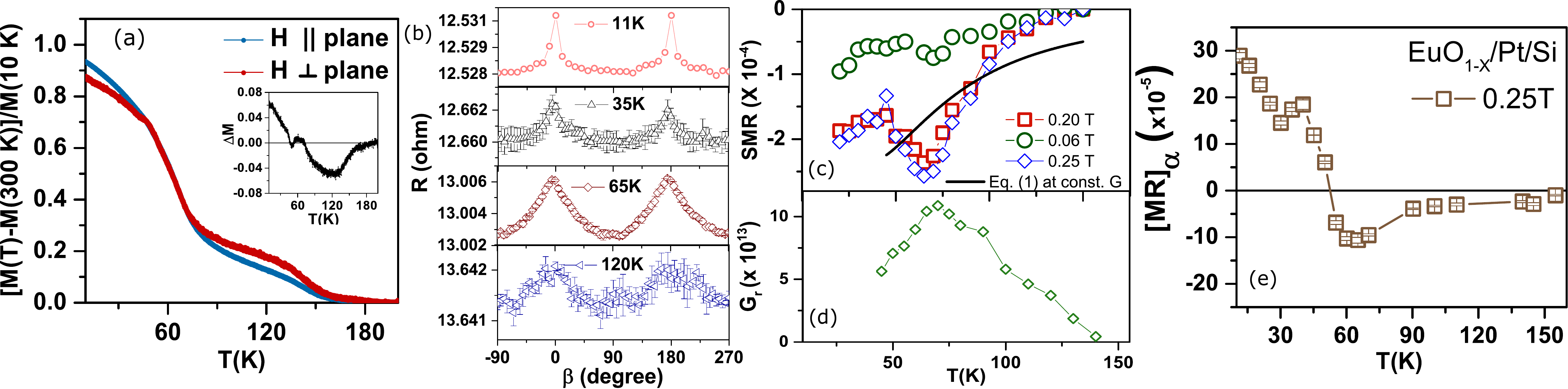}
		\small{\caption{(a) Normalized $M vs T$ measured at FC protocol with 0.25 T $H$ oriented parallel and perpendicular to the sample plane. Inset shows difference of the two curves. (b) ADMR results for rotation along \textit{oopj}. (c) \textit{T}-dependence of MR at different \textit{H}. Black line denotes fit to Equation (1). (d) Calculated G$_{r}$(\textit{T}) obtained by inverting Equation (1). (e) \textit{T}-dependence of AMR in films with the top Pt completely etched.\label{fig:oopj}}}
	\end{center}	
\end{figure*}

Although a negative AMR is opposite to the conventional behavior of ferromagnetic 3\textit{d} alloys\cite{mcguire1975anisotropic}, it has been reported in different systems\cite{kumar2019magnetotransport,alagoz2015mechanism,miyakozawa2016temperature}.
In congruence to these reports, we propose that the observed \textit{T}-dependence of AMR stems from a competition between ip and oop magnetic anisotropy culminating in a complete spin reorientation from oop to ip below 45 K. In order to support our arguments we recorded the magnetization of the sample as a function of \textit{T} after cooling the sample at 0.25 T external field, applied parallel and perpendicular to the sample plane. We represent the scaled magnetization as a function of \textit{T} after subtracting its value at 300 K (that takes into account small offsets) in Fig. \ref{fig:oopj}(a). Firstly, from the figure it is evident that at certain \textit{T}-regions both the ip and oop curves match indicating absence of any preferred orientation in this range (highlighted by the nearly zero $\Delta M$ region in the inset). 
In contrast, for \textit{T}$<$45 K, $M(H\parallel$n) $>$ $M(H\perp$n) i.e. it is relatively easier to orient $M$ ip than oop, whereas for \textit{T}$> \sim$ 69 K, $M(H\parallel$n) $<$ $M(H\perp$n) i.e. the easy axis of orientation is oop. The low-\textit{T} behavior was discussed before as well by illustrating the nature of hysteresis loop for ip and oop $H$ orientations (see Fig. \ref{fig:xrd-vsm}(c)). Such a reorientation of the easy axis with decrease in \textit{T} have been reported before\cite{kumar2019magnetotransport,alagoz2015mechanism,miyakozawa2016temperature}. Incidentally, the observation of increasing trend in AMR below 65 K also coincides with the emergence of exchange dominated ferromagnetic region. As the saturation magnetization increases with decreasing\textit{T}, so does the dipolar shape anisotropy of the thin film, leading to a progressively ip allignment of the easy axis. Once this allignment is complete below 45 K, the AMR changes. However, the oop allignment of the easy axis above 65 K still needs to be elucidated. It is well known that strains in magnetic thin films affect the direction of its anisotropy\cite{alagoz2015mechanism}. Analogously, we argue that the (001) preferential oop orientation of the grains corroborated from x-ray diffraction measurements\cite{mallick2019role} determines the anisotropy direction above 65 K. The sign of the AMR in both these cases can be understood noting that the cross-section for electron scattering is lower when $M$ is perpendicular to the current than when $M$ is parallel i.e. $R(M\parallel J) > R(M\perp J)$, thereby making AMR a useful tool to capture the relative changes in the magnetization components along ip and oop. The small yet non-zero positive AMR above T$_P$ might originate due to induced polarization in the vicinity of T$_P$ due to presence of an applied field. The \textit{H} dependence of high-\textit{T} sign crossover (see supplementary Figure 2 (c)) indeed suggests such a picture. However, further investigation is required to confirm.


Next we demonstrate the presence of an MR from FDMR and ADMR experiments in the oopj geometry as well, within the same \textit{T} and \textit{H} range, which in this case is sensitive to SMR in such HM/FM bilayers. 
The main results are summarized in Fig. \ref{fig:oopj}. In the $R(H)$ results we observe similar hysteretic switching behavior, discernible at low-\textit{T} (see supplementary Figure 3). Noteworthy, we do not observe any change in $R(H)$ above T$_P$. This becomes evident when we plot the MR ratio, defined as $(R(H||t) - R(H||n))/R(H||n)=\Delta R_{MR}/R$, for two \textit{H} values as a function of \textit{T} (see Fig. \ref{fig:oopj}(c)). The negative MR is by convention and corresponds to the "normal" situation\cite{kim2016spin} in the case of pure SMR. The ADMR results illustrated in Fig. \ref{fig:oopj}(b) at 0.25 T, also reflect identical behavior. Notable differences from AMR can be identified, such as the invariance of sign and importantly, disappearance of signal above T$_P$, thereby highlighting difference in physical origin between the two MRs.

Before interpreting the origin of this MR, it is important to consider the contribution from other known MR effects in this geometry, particularly arising due to the presence of texture in the films. Here we will consider two widely known effects; first is the Geometrical Size Effect (GSE) which is a consequence of the crystallinity and anisotropic orientation of the grains\cite{philippi2019impact,kobs2016disentangling}. It is has been shown to be independent of the FM thickness\cite{kobs2011anisotropic} and presumably \textit{T}-independent\cite{gil2005magnetoresistance} as well. Although its behavior near FM's $T_c$ is not well established. Additionally, for GSE dominated ADMR, the maximas appear at $\pm$90$\deg$ which is in contrast to the present data. The second effect known as the Anisotropic Interface Magnetoresistance (AIMR) have been shown to dominate the oopj MR behavior in very thin films, diminishing in magnitude at higher thicknesses\cite{kobs2011anisotropic,kobs2016disentangling,philippi2019impact}. In addition, the same reports found the magnitude to be nearly 2-3 orders smaller in the ip configuration compared to the oop. Subsequently, we expect only AMR contribution to be present in the MR extracted from ip ADMR measurements for a film with the top Pt etched, as shown in Fig. \ref{fig:oopj}(e). Evidently, it reproduces the features similar to the AMR in Fig.\ref{fig:oopt}(b), thereby confirming that the observed MR in oopt for the Pt capped film is dominated by AMR. Furthermore, if significant AIMR is present, then the sum of MRs in the oopt and oopj at different \textit{T} would not be able to corroborate the ip \textit{T}-dependence. Fig. 5 in supplementary suggests that except at very low temperature ($<$45 K i.e when anisotropy is in-plane), the MR extracted from ip ADMR measurements is fairly reproduced by the sum of oopt and oopj MRs (for $H = 0.25 T$).
Consequently, the \textit{T}-dependence of the MR is taken to be pure SMR(\textit{T}) above 45 K. Other than a small AIMR, a proximity induced ferromagnetism in Pt can also result in the low-\textit{T} mismatch, however further experiments are needed to resolve this. Here we focus on the SMR dominated \textit{T}-window, and perform further analysis by adopting the SMR model for metallic bilayers\cite{kim2016spin}, where we renounce the term which accounts for the absorption of the longitudinal spin current by the FM metal since we estimate it to give a negligible contribution. The final form of the equation stands as follows:

  \begin{equation}
  	\frac{\Delta R_{SMR}}{R} = -\theta_{sh}^{2}\frac{\lambda_{Pt}}{t_{Pt}}\frac{tanh^{2}\big(\frac{t_{Pt}}{2\lambda_{Pt}}\big)}{1+\xi}\bigg[\frac{g_r}{1+g_rcoth\big(\frac{t_{Pt}}{\lambda_{Pt}}\big)}\bigg]. 
 \end{equation}
where $g_r=\rho_{Pt} \lambda_{Pt} G_{r}$ and $\xi = \frac{\rho_{Pt} t_{EuO}}{\rho_{EuO} t_{Pt}}$.
Additionally, $\theta_{sh}$ is the spin Hall angle in Pt, $G_{r}$ is the real part of G$_{mix}$ and $t_{Pt(EuO)}$, $\rho_{Pt(EuO)}$, $\lambda_{Pt(EuO)}$ denote the thickness, resistivity and spin diffusion length of Pt(EuO$_{1-x}$) respectively. The \textit{T}-dependent parameters in this equation, particularly G$_{r}$, have been studied extensively by different groups\cite{isasa2015temperature,wang2015spin}. Let us first consider a constant value for $G_{r} = 1.1\times10^{14}$ and try to fit the \textit{T}-dependence of SMR (say 0.20 T data) assuming $\lambda_{Pt}$ to follow a Elliot-Yafet mechanism for spin relaxation, $\lambda_{Pt}=C/T$\cite{vzutic2004spintronics}, while taking $C$ and $\theta_{sh}$ as the fitting parameters. We use the measured $\rho_{Pt}$ and estimated $\rho_{EuO}$) values of resitivity for the fitting. Evidently, as seen in Fig. \ref{fig:oopj}(c), other than suggesting an increase at low-\textit{T}, it fails to reproduce the other features.

Alternately, we can invert Equation (1) to calculate $G_{r}$ such that it reproduces the experimental SMR values exactly. The \textit{T}-dependence of $G_{r}$, calculated from 0.20 T data, with $\lambda_{Pt}=1.4\times10^-7/T$ and $\theta_{sh}=0.057$ (extracted from previous fit but close to reported values\cite{wang2015spin}), is shown in Fig. \ref{fig:oopj}(d). The gradual decay of G$_{r}$ while approaching the Curie temperature has been experimentally demonstrated \cite{wang2015spin, mallick2019role}, which can explain the behavior in the \textit{T}-range between T$_{EuO}$ and T$_P$. However, the decrease in G$_{r}$ below T$_{EuO}$ is still unresolved. Following arguments by Zhou \textit{et al.}\cite{zhou2018spin}, this suppression may be a result of enhanced spin-orbit coupling (SOC), which is not accounted for in SMR theories. Such a scenario is supported by recent observations of enhanced SOC at the Pt/EuO$_{1-x}$ interface\cite{rimal2018interface}. 

In conclusion, we have carried out ADMR and FDMR measurements along three different orientations, from 15 K to above the ferromagnetic transition temperature in Pt/EuO$_{1-x}$. We evaluated AMR and SMR, which constitute important parameters for spintronic applications, from oopt and oopj geometries respectively. In contrast to SMR, the AMR was found to be non-zero even above the ferromagnetic transition temperature, T$_P$, undergoing two sign crossovers in the measured \textit{T}-range, with the low-\textit{T} sign crossover originating from a change in the easy axis of magnetization direction from oop to ip. Finally, we identify a \textit{T}-range wherein the oopj MR is dominated by SMR and implement the theoretical model for SMR to interpret its \textit{T}-dependence based on the spin mixing conductance at the Pt/EuO$_{1-x}$ interface.

\begin{thebibliography}{21}%
\makeatletter
\providecommand \@ifxundefined [1]{%
 \@ifx{#1\undefined}
}%
\providecommand \@ifnum [1]{%
 \ifnum #1\expandafter \@firstoftwo
 \else \expandafter \@secondoftwo
 \fi
}%
\providecommand \@ifx [1]{%
 \ifx #1\expandafter \@firstoftwo
 \else \expandafter \@secondoftwo
 \fi
}%
\providecommand \natexlab [1]{#1}%
\providecommand \enquote  [1]{``#1''}%
\providecommand \bibnamefont  [1]{#1}%
\providecommand \bibfnamefont [1]{#1}%
\providecommand \citenamefont [1]{#1}%
\providecommand \href@noop [0]{\@secondoftwo}%
\providecommand \href [0]{\begingroup \@sanitize@url \@href}%
\providecommand \@href[1]{\@@startlink{#1}\@@href}%
\providecommand \@@href[1]{\endgroup#1\@@endlink}%
\providecommand \@sanitize@url [0]{\catcode `\\12\catcode `\$12\catcode
  `\&12\catcode `\#12\catcode `\^12\catcode `\_12\catcode `\%12\relax}%
\providecommand \@@startlink[1]{}%
\providecommand \@@endlink[0]{}%
\providecommand \url  [0]{\begingroup\@sanitize@url \@url }%
\providecommand \@url [1]{\endgroup\@href {#1}{\urlprefix }}%
\providecommand \urlprefix  [0]{URL }%
\providecommand \Eprint [0]{\href }%
\providecommand \doibase [0]{http://dx.doi.org/}%
\providecommand \selectlanguage [0]{\@gobble}%
\providecommand \bibinfo  [0]{\@secondoftwo}%
\providecommand \bibfield  [0]{\@secondoftwo}%
\providecommand \translation [1]{[#1]}%
\providecommand \BibitemOpen [0]{}%
\providecommand \bibitemStop [0]{}%
\providecommand \bibitemNoStop [0]{.\EOS\space}%
\providecommand \EOS [0]{\spacefactor3000\relax}%
\providecommand \BibitemShut  [1]{\csname bibitem#1\endcsname}%
\let\auto@bib@innerbib\@empty
\bibitem [{\citenamefont {Passell}, \citenamefont {Dietrich},\ and\
  \citenamefont {Als-Nielsen}(1976)}]{passell1976neutron}%
  \BibitemOpen
  \bibfield  {author} {\bibinfo {author} {\bibfnamefont {L.}~\bibnamefont
  {Passell}}, \bibinfo {author} {\bibfnamefont {O.}~\bibnamefont {Dietrich}}, \
  and\ \bibinfo {author} {\bibfnamefont {J.}~\bibnamefont {Als-Nielsen}},\
  }\bibfield  {title} {\enquote {\bibinfo {title} {Neutron scattering from the
  heisenberg ferromagnets euo and eus. i. the exchange interactions},}\
  }\href@noop {} {\bibfield  {journal} {\bibinfo  {journal} {Physical Review
  B}\ }\textbf {\bibinfo {volume} {14}},\ \bibinfo {pages} {4897} (\bibinfo
  {year} {1976})}\BibitemShut {NoStop}%
\bibitem [{\citenamefont {Steeneken}\ \emph {et~al.}(2002)\citenamefont
  {Steeneken}, \citenamefont {Tjeng}, \citenamefont {Elfimov}, \citenamefont
  {Sawatzky}, \citenamefont {Ghiringhelli}, \citenamefont {Brookes},\ and\
  \citenamefont {Huang}}]{steeneken2002exchange}%
  \BibitemOpen
  \bibfield  {author} {\bibinfo {author} {\bibfnamefont {P.}~\bibnamefont
  {Steeneken}}, \bibinfo {author} {\bibfnamefont {L.}~\bibnamefont {Tjeng}},
  \bibinfo {author} {\bibfnamefont {I.}~\bibnamefont {Elfimov}}, \bibinfo
  {author} {\bibfnamefont {G.}~\bibnamefont {Sawatzky}}, \bibinfo {author}
  {\bibfnamefont {G.}~\bibnamefont {Ghiringhelli}}, \bibinfo {author}
  {\bibfnamefont {N.}~\bibnamefont {Brookes}}, \ and\ \bibinfo {author}
  {\bibfnamefont {D.-J.}\ \bibnamefont {Huang}},\ }\bibfield  {title} {\enquote
  {\bibinfo {title} {Exchange splitting and charge carrier spin polarization in
  euo},}\ }\href@noop {} {\bibfield  {journal} {\bibinfo  {journal} {Physical
  review letters}\ }\textbf {\bibinfo {volume} {88}},\ \bibinfo {pages}
  {047201} (\bibinfo {year} {2002})}\BibitemShut {NoStop}%
\bibitem [{\citenamefont {Rimal}\ and\ \citenamefont
  {Tang}(2018)}]{rimal2018interface}%
  \BibitemOpen
  \bibfield  {author} {\bibinfo {author} {\bibfnamefont {G.}~\bibnamefont
  {Rimal}}\ and\ \bibinfo {author} {\bibfnamefont {J.}~\bibnamefont {Tang}},\
  }\bibfield  {title} {\enquote {\bibinfo {title} {Interface enhanced magnetic
  anisotropy in pt/euo 1- x films},}\ }\href@noop {} {\bibfield  {journal}
  {\bibinfo  {journal} {Physical Review B}\ }\textbf {\bibinfo {volume} {98}},\
  \bibinfo {pages} {144442} (\bibinfo {year} {2018})}\BibitemShut {NoStop}%
\bibitem [{\citenamefont {Althammer}\ \emph {et~al.}(2013)\citenamefont
  {Althammer}, \citenamefont {Meyer}, \citenamefont {Nakayama}, \citenamefont
  {Schreier}, \citenamefont {Altmannshofer}, \citenamefont {Weiler},
  \citenamefont {Huebl}, \citenamefont {Gepr{\"a}gs}, \citenamefont {Opel},
  \citenamefont {Gross} \emph {et~al.}}]{althammer2013quantitative}%
  \BibitemOpen
  \bibfield  {author} {\bibinfo {author} {\bibfnamefont {M.}~\bibnamefont
  {Althammer}}, \bibinfo {author} {\bibfnamefont {S.}~\bibnamefont {Meyer}},
  \bibinfo {author} {\bibfnamefont {H.}~\bibnamefont {Nakayama}}, \bibinfo
  {author} {\bibfnamefont {M.}~\bibnamefont {Schreier}}, \bibinfo {author}
  {\bibfnamefont {S.}~\bibnamefont {Altmannshofer}}, \bibinfo {author}
  {\bibfnamefont {M.}~\bibnamefont {Weiler}}, \bibinfo {author} {\bibfnamefont
  {H.}~\bibnamefont {Huebl}}, \bibinfo {author} {\bibfnamefont
  {S.}~\bibnamefont {Gepr{\"a}gs}}, \bibinfo {author} {\bibfnamefont
  {M.}~\bibnamefont {Opel}}, \bibinfo {author} {\bibfnamefont {R.}~\bibnamefont
  {Gross}},  \emph {et~al.},\ }\bibfield  {title} {\enquote {\bibinfo {title}
  {Quantitative study of the spin hall magnetoresistance in ferromagnetic
  insulator/normal metal hybrids},}\ }\href@noop {} {\bibfield  {journal}
  {\bibinfo  {journal} {Physical Review B}\ }\textbf {\bibinfo {volume} {87}},\
  \bibinfo {pages} {224401} (\bibinfo {year} {2013})}\BibitemShut {NoStop}%
\bibitem [{\citenamefont {Wang}\ \emph {et~al.}(2015)\citenamefont {Wang},
  \citenamefont {Zou}, \citenamefont {Zhang}, \citenamefont {Cai},
  \citenamefont {Wang}, \citenamefont {Shen},\ and\ \citenamefont
  {Sun}}]{wang2015spin}%
  \BibitemOpen
  \bibfield  {author} {\bibinfo {author} {\bibfnamefont {S.}~\bibnamefont
  {Wang}}, \bibinfo {author} {\bibfnamefont {L.}~\bibnamefont {Zou}}, \bibinfo
  {author} {\bibfnamefont {X.}~\bibnamefont {Zhang}}, \bibinfo {author}
  {\bibfnamefont {J.}~\bibnamefont {Cai}}, \bibinfo {author} {\bibfnamefont
  {S.}~\bibnamefont {Wang}}, \bibinfo {author} {\bibfnamefont {B.}~\bibnamefont
  {Shen}}, \ and\ \bibinfo {author} {\bibfnamefont {J.}~\bibnamefont {Sun}},\
  }\bibfield  {title} {\enquote {\bibinfo {title} {Spin seebeck effect and spin
  hall magnetoresistance at high temperatures for a pt/yttrium iron garnet
  hybrid structure},}\ }\href@noop {} {\bibfield  {journal} {\bibinfo
  {journal} {Nanoscale}\ }\textbf {\bibinfo {volume} {7}},\ \bibinfo {pages}
  {17812--17819} (\bibinfo {year} {2015})}\BibitemShut {NoStop}%
\bibitem [{\citenamefont {Mallick}, \citenamefont {Wagh},\ and\ \citenamefont
  {Kumar}(2019)}]{mallick2019enhanced}%
  \BibitemOpen
  \bibfield  {author} {\bibinfo {author} {\bibfnamefont {K.}~\bibnamefont
  {Mallick}}, \bibinfo {author} {\bibfnamefont {A.~A.}\ \bibnamefont {Wagh}}, \
  and\ \bibinfo {author} {\bibfnamefont {P.~A.}\ \bibnamefont {Kumar}},\
  }\bibfield  {title} {\enquote {\bibinfo {title} {Enhanced spin transport in a
  ferrite having distributed energy barriers for exchange bias},}\ }\href@noop
  {} {\bibfield  {journal} {\bibinfo  {journal} {Journal of Magnetism and
  Magnetic Materials}\ }\textbf {\bibinfo {volume} {492}},\ \bibinfo {pages}
  {165644} (\bibinfo {year} {2019})}\BibitemShut {NoStop}%
\bibitem [{\citenamefont {Mallick}\ \emph {et~al.}(2019)\citenamefont
  {Mallick}, \citenamefont {Wagh}, \citenamefont {Ionescu}, \citenamefont
  {Barnes},\ and\ \citenamefont {Kumar}}]{mallick2019role}%
  \BibitemOpen
  \bibfield  {author} {\bibinfo {author} {\bibfnamefont {K.}~\bibnamefont
  {Mallick}}, \bibinfo {author} {\bibfnamefont {A.~A.}\ \bibnamefont {Wagh}},
  \bibinfo {author} {\bibfnamefont {A.}~\bibnamefont {Ionescu}}, \bibinfo
  {author} {\bibfnamefont {C.~H.}\ \bibnamefont {Barnes}}, \ and\ \bibinfo
  {author} {\bibfnamefont {P.~A.}\ \bibnamefont {Kumar}},\ }\bibfield  {title}
  {\enquote {\bibinfo {title} {Role of spin mixing conductance in determining
  thermal spin pumping near the ferromagnetic phase transition in euo 1- x and
  la 2 nimno 6},}\ }\href@noop {} {\bibfield  {journal} {\bibinfo  {journal}
  {Physical Review B}\ }\textbf {\bibinfo {volume} {100}},\ \bibinfo {pages}
  {224403} (\bibinfo {year} {2019})}\BibitemShut {NoStop}%
\bibitem [{\citenamefont {Kim}\ \emph {et~al.}(2016)\citenamefont {Kim},
  \citenamefont {Sheng}, \citenamefont {Takahashi}, \citenamefont {Mitani},\
  and\ \citenamefont {Hayashi}}]{kim2016spin}%
  \BibitemOpen
  \bibfield  {author} {\bibinfo {author} {\bibfnamefont {J.}~\bibnamefont
  {Kim}}, \bibinfo {author} {\bibfnamefont {P.}~\bibnamefont {Sheng}}, \bibinfo
  {author} {\bibfnamefont {S.}~\bibnamefont {Takahashi}}, \bibinfo {author}
  {\bibfnamefont {S.}~\bibnamefont {Mitani}}, \ and\ \bibinfo {author}
  {\bibfnamefont {M.}~\bibnamefont {Hayashi}},\ }\bibfield  {title} {\enquote
  {\bibinfo {title} {Spin hall magnetoresistance in metallic bilayers},}\
  }\href@noop {} {\bibfield  {journal} {\bibinfo  {journal} {Physical review
  letters}\ }\textbf {\bibinfo {volume} {116}},\ \bibinfo {pages} {097201}
  (\bibinfo {year} {2016})}\BibitemShut {NoStop}%
\bibitem [{\citenamefont {Zhou}\ \emph {et~al.}(2018)\citenamefont {Zhou},
  \citenamefont {Seki}, \citenamefont {Kubota}, \citenamefont {Bauer},\ and\
  \citenamefont {Takanashi}}]{zhou2018spin}%
  \BibitemOpen
  \bibfield  {author} {\bibinfo {author} {\bibfnamefont {W.}~\bibnamefont
  {Zhou}}, \bibinfo {author} {\bibfnamefont {T.}~\bibnamefont {Seki}}, \bibinfo
  {author} {\bibfnamefont {T.}~\bibnamefont {Kubota}}, \bibinfo {author}
  {\bibfnamefont {G.}~\bibnamefont {Bauer}}, \ and\ \bibinfo {author}
  {\bibfnamefont {K.}~\bibnamefont {Takanashi}},\ }\bibfield  {title} {\enquote
  {\bibinfo {title} {Spin-hall and anisotropic magnetoresistance in
  ferrimagnetic co-gd/pt layers},}\ }\href@noop {} {\bibfield  {journal}
  {\bibinfo  {journal} {Physical Review Materials}\ }\textbf {\bibinfo {volume}
  {2}},\ \bibinfo {pages} {094404} (\bibinfo {year} {2018})}\BibitemShut
  {NoStop}%
\bibitem [{\citenamefont {McGuire}\ and\ \citenamefont
  {Potter}(1975)}]{mcguire1975anisotropic}%
  \BibitemOpen
  \bibfield  {author} {\bibinfo {author} {\bibfnamefont {T.}~\bibnamefont
  {McGuire}}\ and\ \bibinfo {author} {\bibfnamefont {R.}~\bibnamefont
  {Potter}},\ }\bibfield  {title} {\enquote {\bibinfo {title} {Anisotropic
  magnetoresistance in ferromagnetic 3d alloys},}\ }\href@noop {} {\bibfield
  {journal} {\bibinfo  {journal} {IEEE Transactions on Magnetics}\ }\textbf
  {\bibinfo {volume} {11}},\ \bibinfo {pages} {1018--1038} (\bibinfo {year}
  {1975})}\BibitemShut {NoStop}%
\bibitem [{\citenamefont {Alagoz}\ \emph {et~al.}(2015)\citenamefont {Alagoz},
  \citenamefont {Desomberg}, \citenamefont {Taheri}, \citenamefont {Razavi},
  \citenamefont {Chow},\ and\ \citenamefont {Jung}}]{alagoz2015mechanism}%
  \BibitemOpen
  \bibfield  {author} {\bibinfo {author} {\bibfnamefont {H.}~\bibnamefont
  {Alagoz}}, \bibinfo {author} {\bibfnamefont {J.}~\bibnamefont {Desomberg}},
  \bibinfo {author} {\bibfnamefont {M.}~\bibnamefont {Taheri}}, \bibinfo
  {author} {\bibfnamefont {F.}~\bibnamefont {Razavi}}, \bibinfo {author}
  {\bibfnamefont {K.}~\bibnamefont {Chow}}, \ and\ \bibinfo {author}
  {\bibfnamefont {J.}~\bibnamefont {Jung}},\ }\bibfield  {title} {\enquote
  {\bibinfo {title} {Mechanism of sign crossover of the anisotropic
  magneto-resistance in la0. 7- x pr x ca0. 3mno3 thin films},}\ }\href@noop {}
  {\bibfield  {journal} {\bibinfo  {journal} {Applied Physics Letters}\
  }\textbf {\bibinfo {volume} {106}},\ \bibinfo {pages} {082407} (\bibinfo
  {year} {2015})}\BibitemShut {NoStop}%
\bibitem [{\citenamefont {Kumar}\ \emph {et~al.}(2019)\citenamefont {Kumar},
  \citenamefont {Soh}, \citenamefont {Wang},\ and\ \citenamefont
  {Xiong}}]{kumar2019magnetotransport}%
  \BibitemOpen
  \bibfield  {author} {\bibinfo {author} {\bibfnamefont {N.}~\bibnamefont
  {Kumar}}, \bibinfo {author} {\bibfnamefont {Y.}~\bibnamefont {Soh}}, \bibinfo
  {author} {\bibfnamefont {Y.}~\bibnamefont {Wang}}, \ and\ \bibinfo {author}
  {\bibfnamefont {Y.}~\bibnamefont {Xiong}},\ }\bibfield  {title} {\enquote
  {\bibinfo {title} {Magnetotransport as a diagnostic of spin reorientation:
  Kagome ferromagnet as a case study},}\ }\href@noop {} {\bibfield  {journal}
  {\bibinfo  {journal} {Physical Review B}\ }\textbf {\bibinfo {volume}
  {100}},\ \bibinfo {pages} {214420} (\bibinfo {year} {2019})}\BibitemShut
  {NoStop}%
\bibitem [{\citenamefont {Miyakozawa}\ \emph {et~al.}(2016)\citenamefont
  {Miyakozawa}, \citenamefont {Chen}, \citenamefont {Matsukura},\ and\
  \citenamefont {Ohno}}]{miyakozawa2016temperature}%
  \BibitemOpen
  \bibfield  {author} {\bibinfo {author} {\bibfnamefont {S.}~\bibnamefont
  {Miyakozawa}}, \bibinfo {author} {\bibfnamefont {L.}~\bibnamefont {Chen}},
  \bibinfo {author} {\bibfnamefont {F.}~\bibnamefont {Matsukura}}, \ and\
  \bibinfo {author} {\bibfnamefont {H.}~\bibnamefont {Ohno}},\ }\bibfield
  {title} {\enquote {\bibinfo {title} {Temperature dependence of in-plane
  magnetic anisotropy and anisotropic magnetoresistance in (ga, mn) as codoped
  with li},}\ }\href@noop {} {\bibfield  {journal} {\bibinfo  {journal}
  {Applied Physics Letters}\ }\textbf {\bibinfo {volume} {108}},\ \bibinfo
  {pages} {112404} (\bibinfo {year} {2016})}\BibitemShut {NoStop}%
\bibitem [{\citenamefont {Altendorf}\ \emph {et~al.}(2011)\citenamefont
  {Altendorf}, \citenamefont {Efimenko}, \citenamefont {Oliana}, \citenamefont
  {Kierspel}, \citenamefont {Rata},\ and\ \citenamefont
  {Tjeng}}]{altendorf2011oxygen}%
  \BibitemOpen
  \bibfield  {author} {\bibinfo {author} {\bibfnamefont {S.}~\bibnamefont
  {Altendorf}}, \bibinfo {author} {\bibfnamefont {A.}~\bibnamefont {Efimenko}},
  \bibinfo {author} {\bibfnamefont {V.}~\bibnamefont {Oliana}}, \bibinfo
  {author} {\bibfnamefont {H.}~\bibnamefont {Kierspel}}, \bibinfo {author}
  {\bibfnamefont {A.}~\bibnamefont {Rata}}, \ and\ \bibinfo {author}
  {\bibfnamefont {L.}~\bibnamefont {Tjeng}},\ }\bibfield  {title} {\enquote
  {\bibinfo {title} {Oxygen off-stoichiometry and phase separation in euo thin
  films},}\ }\href@noop {} {\bibfield  {journal} {\bibinfo  {journal} {Physical
  Review B}\ }\textbf {\bibinfo {volume} {84}},\ \bibinfo {pages} {155442}
  (\bibinfo {year} {2011})}\BibitemShut {NoStop}%
\bibitem [{\citenamefont {Steeneken}(2012)}]{steeneken2012new}%
  \BibitemOpen
  \bibfield  {author} {\bibinfo {author} {\bibfnamefont {P.~G.}\ \bibnamefont
  {Steeneken}},\ }\bibfield  {title} {\enquote {\bibinfo {title} {New light on
  euo thin films: Preparation, transport, magnetism and spectroscopy of a
  ferromagnetic semiconductor},}\ }\href@noop {} {\bibfield  {journal}
  {\bibinfo  {journal} {arXiv preprint arXiv:1203.6771}\ } (\bibinfo {year}
  {2012})}\BibitemShut {NoStop}%
\bibitem [{\citenamefont {Philippi-Kobs}\ \emph {et~al.}(2019)\citenamefont
  {Philippi-Kobs}, \citenamefont {Farhadi}, \citenamefont {Matheis},
  \citenamefont {Lott}, \citenamefont {Chuvilin},\ and\ \citenamefont
  {Oepen}}]{philippi2019impact}%
  \BibitemOpen
  \bibfield  {author} {\bibinfo {author} {\bibfnamefont {A.}~\bibnamefont
  {Philippi-Kobs}}, \bibinfo {author} {\bibfnamefont {A.}~\bibnamefont
  {Farhadi}}, \bibinfo {author} {\bibfnamefont {L.}~\bibnamefont {Matheis}},
  \bibinfo {author} {\bibfnamefont {D.}~\bibnamefont {Lott}}, \bibinfo {author}
  {\bibfnamefont {A.}~\bibnamefont {Chuvilin}}, \ and\ \bibinfo {author}
  {\bibfnamefont {H.}~\bibnamefont {Oepen}},\ }\bibfield  {title} {\enquote
  {\bibinfo {title} {Impact of symmetry on anisotropic magnetoresistance in
  textured ferromagnetic thin films},}\ }\href@noop {} {\bibfield  {journal}
  {\bibinfo  {journal} {Physical review letters}\ }\textbf {\bibinfo {volume}
  {123}},\ \bibinfo {pages} {137201} (\bibinfo {year} {2019})}\BibitemShut
  {NoStop}%
\bibitem [{\citenamefont {Kobs}\ and\ \citenamefont
  {Oepen}(2016)}]{kobs2016disentangling}%
  \BibitemOpen
  \bibfield  {author} {\bibinfo {author} {\bibfnamefont {A.}~\bibnamefont
  {Kobs}}\ and\ \bibinfo {author} {\bibfnamefont {H.~P.}\ \bibnamefont
  {Oepen}},\ }\bibfield  {title} {\enquote {\bibinfo {title} {Disentangling
  interface and bulk contributions to the anisotropic magnetoresistance in
  pt/co/pt sandwiches},}\ }\href@noop {} {\bibfield  {journal} {\bibinfo
  {journal} {Physical Review B}\ }\textbf {\bibinfo {volume} {93}},\ \bibinfo
  {pages} {014426} (\bibinfo {year} {2016})}\BibitemShut {NoStop}%
\bibitem [{\citenamefont {Kobs}\ \emph {et~al.}(2011)\citenamefont {Kobs},
  \citenamefont {He{\ss}e}, \citenamefont {Kreuzpaintner}, \citenamefont
  {Winkler}, \citenamefont {Lott}, \citenamefont {Weinberger}, \citenamefont
  {Schreyer},\ and\ \citenamefont {Oepen}}]{kobs2011anisotropic}%
  \BibitemOpen
  \bibfield  {author} {\bibinfo {author} {\bibfnamefont {A.}~\bibnamefont
  {Kobs}}, \bibinfo {author} {\bibfnamefont {S.}~\bibnamefont {He{\ss}e}},
  \bibinfo {author} {\bibfnamefont {W.}~\bibnamefont {Kreuzpaintner}}, \bibinfo
  {author} {\bibfnamefont {G.}~\bibnamefont {Winkler}}, \bibinfo {author}
  {\bibfnamefont {D.}~\bibnamefont {Lott}}, \bibinfo {author} {\bibfnamefont
  {P.}~\bibnamefont {Weinberger}}, \bibinfo {author} {\bibfnamefont
  {A.}~\bibnamefont {Schreyer}}, \ and\ \bibinfo {author} {\bibfnamefont
  {H.}~\bibnamefont {Oepen}},\ }\bibfield  {title} {\enquote {\bibinfo {title}
  {Anisotropic interface magnetoresistance in pt/co/pt sandwiches},}\
  }\href@noop {} {\bibfield  {journal} {\bibinfo  {journal} {Physical review
  letters}\ }\textbf {\bibinfo {volume} {106}},\ \bibinfo {pages} {217207}
  (\bibinfo {year} {2011})}\BibitemShut {NoStop}%
\bibitem [{\citenamefont {Gil}\ \emph {et~al.}(2005)\citenamefont {Gil},
  \citenamefont {G{\"o}rlitz}, \citenamefont {Horisberger},\ and\ \citenamefont
  {K{\"o}tzler}}]{gil2005magnetoresistance}%
  \BibitemOpen
  \bibfield  {author} {\bibinfo {author} {\bibfnamefont {W.}~\bibnamefont
  {Gil}}, \bibinfo {author} {\bibfnamefont {D.}~\bibnamefont {G{\"o}rlitz}},
  \bibinfo {author} {\bibfnamefont {M.}~\bibnamefont {Horisberger}}, \ and\
  \bibinfo {author} {\bibfnamefont {J.}~\bibnamefont {K{\"o}tzler}},\
  }\bibfield  {title} {\enquote {\bibinfo {title} {Magnetoresistance anisotropy
  of polycrystalline cobalt films: Geometrical-size and domain effects},}\
  }\href@noop {} {\bibfield  {journal} {\bibinfo  {journal} {Physical Review
  B}\ }\textbf {\bibinfo {volume} {72}},\ \bibinfo {pages} {134401} (\bibinfo
  {year} {2005})}\BibitemShut {NoStop}%
\bibitem [{\citenamefont {Isasa}\ \emph {et~al.}(2015)\citenamefont {Isasa},
  \citenamefont {Villamor}, \citenamefont {Hueso}, \citenamefont {Gradhand},\
  and\ \citenamefont {Casanova}}]{isasa2015temperature}%
  \BibitemOpen
  \bibfield  {author} {\bibinfo {author} {\bibfnamefont {M.}~\bibnamefont
  {Isasa}}, \bibinfo {author} {\bibfnamefont {E.}~\bibnamefont {Villamor}},
  \bibinfo {author} {\bibfnamefont {L.~E.}\ \bibnamefont {Hueso}}, \bibinfo
  {author} {\bibfnamefont {M.}~\bibnamefont {Gradhand}}, \ and\ \bibinfo
  {author} {\bibfnamefont {F.}~\bibnamefont {Casanova}},\ }\bibfield  {title}
  {\enquote {\bibinfo {title} {Temperature dependence of spin diffusion length
  and spin hall angle in au and pt},}\ }\href@noop {} {\bibfield  {journal}
  {\bibinfo  {journal} {Physical Review B}\ }\textbf {\bibinfo {volume} {91}},\
  \bibinfo {pages} {024402} (\bibinfo {year} {2015})}\BibitemShut {NoStop}%
\bibitem [{\citenamefont {{\v{Z}}uti{\'c}}, \citenamefont {Fabian},\ and\
  \citenamefont {Sarma}(2004)}]{vzutic2004spintronics}%
  \BibitemOpen
  \bibfield  {author} {\bibinfo {author} {\bibfnamefont {I.}~\bibnamefont
  {{\v{Z}}uti{\'c}}}, \bibinfo {author} {\bibfnamefont {J.}~\bibnamefont
  {Fabian}}, \ and\ \bibinfo {author} {\bibfnamefont {S.~D.}\ \bibnamefont
  {Sarma}},\ }\bibfield  {title} {\enquote {\bibinfo {title} {Spintronics:
  Fundamentals and applications},}\ }\href@noop {} {\bibfield  {journal}
  {\bibinfo  {journal} {Reviews of modern physics}\ }\textbf {\bibinfo {volume}
  {76}},\ \bibinfo {pages} {323} (\bibinfo {year} {2004})}\BibitemShut
  {NoStop}%
\end{thebibliography}

%

\end{document}